\documentstyle[a4,12pt]{article}
\renewcommand{\baselinestretch}{1.5}

\begin{document}

\begin{flushright}
{\large\bf LMU-13/94} \\
{(Revised version)}
\end{flushright}

\vspace{0.2cm}

\begin{center}
{\Large\bf Remarks on the Quark-diagram Description \\
of Two-body Nonleptonic $B$-meson Decays}
\end{center}

\vspace{1.2cm}

\begin{center}
{\bf Zhi-zhong XING}\footnote{
E-mail: Xing$@$hep.physik.uni-muenchen.de }\\
{\sl Sektion Physik, Theoretische Physik, Universit${\sl\ddot a}$t M${\sl\ddot
u}$nchen,}\\
{\sl Theresienstrasse 37, D-80333 Munich, Germany}
\end{center}

\vspace{2cm}

\begin{abstract}

	To lowest-order weak interactions in the standard model, we point out that
a complete graph description of two-body mesonic $B$ (or $D$) decays needs ten
topologically different quark diagrams.
The two-body baryonic $B$ decays can be illustrated in terms of five diagrams.
We remark a variety of features of the graph language. Some pure channels of
mesonic
$B$ decays, which can be used to test the quark-diagram scheme, are discussed
in some detail.

\end{abstract}

\vspace{3cm}

\begin{center}
PACS number(s): $~$ 14.40.J, 13.25
\end{center}

\newpage

	The nonleptonic weak decays of $B$ mesons appear to be a valuable window for
determining the quark mixing parameters, probing the origin of $CP$ violation,
and investigating the
nonperturbative confinement forces. To date, some experimental data on many
specific channels of this
nature have been accumulated [1]. In the framework of the standard model,
the dynamics of exclusive nonleptonic decays is not yet well known. Hence one
has to
rely, in most cases, on approximate methods or models for quantitative studies.
In the literature, the quark-diagram language has been extensively applied to
the phenomenology of weak
$B$ transitions [2,3]. It proves a simple and intuitive approach, which enables
one to gain
some salient features of the decays under discussion before carrying out
realistic calculations.

\vspace{0.3cm}

	This short note is to give some new and non-trivial remarks on applications of
the quark-diagram method to
nonleptonic weak decays of $B$ mesons. Improving the six-graph scheme presented
by Chau [3],
we show that a complete description of two-body mesonic $B$ (or $D$)
decays needs {\it ten} topologically different quark diagrams.
The two-body baryonic $B$ decays can be described in terms of five quark
diagrams.
We point out a variety of features of the graph language. Some pure channels of
mesonic $B$ decays,
which can be used to test the quark-diagram scheme, are discussed in some
detail.
We emphasize that a comparison between theoretical predictions and experimental
data on the single-graph induced $B$ decays could serve as a useful probe of
the dynamics of nonleptonic
weak transitions.

\vspace{0.3cm}

	To lowest-order weak interactions in the standard model, the nonleptonic weak
decays are governed
by a single $W$-exchange current with flavour changes. It is known that the
inclusive nonleptonic decays
of $B$ (or $D$) mesons can be illustrated by six topologically distinct quark
diagrams [3]. For the exclusive
case, the quarks and antiquarks occurring through $\Delta B=\pm 1$ (or $\Delta
C=\pm 1$) transitions are combined
by final-state strong interactions into specific hadron (meson or baryon)
states.
To describe these hadrons in a pictorial approach, the quark-antiquark pairs
need to be created from the vacuum
for some inclusive graphs. This ensures that the final-state valence quarks can
assemble and hadronize
themselves in all possible ways to form a two-body or multi-body product [2,3].
In this picture,
one has assumed that all strong-interaction effects
(in the form of all possible gluon lines) are included in the quark diagrams.

\vspace{0.3cm}

	Based on the above arguments, we present in Fig. 1 a set of ten topologically
different quark diagrams for a complete description of two-body mesonic $B$ (or
$D$) decays.
The charged mesons $B^{\pm}_{u}, B^{\pm}_{c}, D^{\pm}$, and $D^{\pm}_{s}$ can
decay via the
graphs 1 ($1^{'}$), 3 ($3^{'}$), and 4 ($4^{'}$); while the neutral mesons
$\stackrel{(-)}{B}$$^{0}_{d}, \stackrel{(-)}{B}$$^{0}_{s}$,
and $\stackrel{(-)}{D}$$^{0}$ can decay through the graphs 1 ($1^{'}$), 2
($2^{'}$), 4 ($4^{'}$), and 5 ($5^{'}$).
In a similar way one can present a graph scheme for describing the two-body
baryonic decays of
$B$ mesons. There are five quark diagrams for a $B$ meson decaying to two light
baryons,
as illustrated in Fig. 2. The mesons $B^{\pm}_{u}$ and $B^{\pm}_{c}$ can decay
through the
graphs (a), (c), and (d); while $\stackrel{(-)}{B}$$^{0}_{d}$ and
$\stackrel{(-)}
{B}$$^{0}_{s}$ can decay via the graphs (a), (b), (d), and (e).
Some useful remarks on the above graph scheme are in order.

\vspace{0.3cm}

	(i) The diagrams $n$ and $n'$ ($n=1,2,3,4$, or 5) are different from each
other in the
final-state hadronization of valence quarks. One of them is colour-suppressed.
It should be
noted that the graphs $2^{'} - 4^{'}$ apply only to the decays where one
final-state meson is a flavour
singlet, and that $5^{'}$ applies only to the processes where both final-state
mesons are flavour singlets.

\vspace{0.3cm}

	(ii) Improving the six-graph scheme given previously by Chau [3], here we have
taken four
additional diagrams ($2^{'}, 3^{'}, 4^{'}$, and $5^{'}$) into account. Our
present scheme can give a
complete graph description of two-body mesonic $B$ (or $D$) decays. In fact,
there are not enough dynamic
or empiric grounds to justify that contributions from the graphs $2^{'}-4^{'}$
should be negligibly
smaller than those from the corresponding graphs $2-4$.

\vspace{0.3cm}

	(iii) The diagrams 4 ($4^{'}$) and 5 ($5^{'}$) are the loop-induced penguin
transitions.
The quark-antiquark pairs in these graphs can be produced via both strong (hard
and soft gluons)
and electroweak ($\gamma, Z^{0}$, and $H^{0}$) interactions. The latter is in
general sensitive to the
flavour of quarks to which the $\gamma$, $Z^{0}$, or $H^{0}$ couples [4].

\vspace{0.3cm}

	(iv) In many practical calculations, the $W$-exchange (2 and $2^{'}$)
and annihilation (3, $3^{'}$ and 5, $5^{'}$) diagrams are argued to be helicity
unfavoured or formfactor suppressed. However, the helicity
suppression is possible to be lifted when soft gluon effects are taken into
account [5]. On the other
hand, the existing evaluation of annihilation formfactors for exclusive $B$ and
$D$ decays has many
uncertainties. It is quite possible that significant final-state rescattering
effects
boost the $W$-exchange or annihilation channels of $D$ and $B$ decays [6].

\vspace{0.3cm}

	(v) It should be noted that $B^{\pm}_{c}$ mesons can decay through either the
$c$-quark spectator graph or the $b$-quark one. The latter case is illustrated
in
Fig. 3, where the light $B$ mesons ($B^{\pm}_{u}, \stackrel{(-)}{B}$$^{0}_{d}$,
and $\stackrel{(-)}{B}$$^{0}_{s}$) are produced via the semileptonic or
nonleptonic $B^{\pm}_{c}$ transitions.

\vspace{0.3cm}

	(vi) All five graphs in Fig. 2 are colour-suppressed. For a colour-singlet
baryon in the final state,
the colour states of its three valence quarks should be different from one
another. This leads to
a colour-suppression factor $2/9$, included in every quark-diagram amplitude
[7].

\vspace{0.3cm}

	The direct way for testing the above graph scheme is to measure the pure decay
modes
induced by a single quark diagram, in which the uncertainty from graph mixing
is
avoidable or relatively small. In Table 1, we list some pure channels of
two-body mesonic
$B$ decays. Clearly the quark diagrams 1$-$3 and $1^{'}-3^{'}$ can be
across-checked by
detecting the respective pure decay modes of $B_{u}, B_{d}, B_{s}$, and $B_{c}$
mesons.
Note that there is no pure channel occurring only through the graph 4 ($4^{'}$)
or 5 ($5^{'}$).
Measuring $\bar{B}^{0}_{d}\rightarrow \bar{K}^{0}\phi$ or
$\bar{B}^{0}_{s}\rightarrow K^{0}\phi$
may test the graphs 4 and $4^{'}$ as a whole. Similarly one can look for the
pure two-body
baryonic decays of $B$ mesons. Unfortunately, it is still a long run to study
baryonic $B$ decays
in a quantitative and systematic way, because the relevant data have not been
accumulated.

\vspace{0.3cm}

	Phenomenologically it is worth while to pay more attention to the pure decay
modes listed in Table 1.
Those channels occurring through the spectator graphs 1 and $1^{'}$ can be
calculated by using the
tree-level effective Hamiltonian [6,8] and the factorization approximation [9].
For illustration, we estimate their branching ratios in the context of the
Bauer-Stech-Wirbel (BSW) model [6].
Our numerical results, accompanied by the current experimental limits to these
processes,
are listed in Table 2. One can observe that the pure $B^{-}_{u}$ and
$\bar{B}^{0}_{d}$ decays via
the graph 1 are measurable in the near future. The decay modes
$\bar{B}^{0}_{s}\rightarrow
D^{(*)+}_{s} + (\pi^{-}, \rho^{-}, a^{-}_{1})$ are also promising for
experimental observation.
In contrast, the transitions via the graph $1^{'}$ have relatively smaller
branching ratios.
It should be emphasized that the pure channels $\bar{B}^{0}_{d}\rightarrow
\stackrel{(-)}{D}$$^{(*)0}
\bar{K}^{(*)0}$ and their $CP$-conjugate processes are of large interest for
studying $CP$
violation and testing the Cabibbo-Kobayashi-Maskawa (CKM) unitarity triangle
[11,12].

\vspace{0.3cm}

	It is difficult to estimate branching ratios of the decay modes via the graphs
2 ($2^{'}$)
and 3 ($3^{'}$), since the relevant annihilation formfactors are not known.
However,
theoretical difficulties cannot obstruct experimental attempts to detect the
pure channels
occurring only through the $W$-exchange or annihilation diagrams. For instance,
the existing data give
Br$(B^{-}_{u}\rightarrow D^{-}_{s}K^{0})< 1.1\times 10^{-3}$ and
Br$(\bar{B}^{0}_{d}\rightarrow
D^{+}_{s}K^{-})<2.4\times 10^{-4}$ [1]. In comparison with the tree-level
transition, those processes via the diagrams 4 ($4^{'}$) and 5 ($5^{'}$) are
more difficult
to control in phenomenology. Using the penguin effective Hamiltonian [13] and
the factorization approximation,
one can roughly estimate the branching ratios of $\bar{B}^{0}_{d}\rightarrow
\bar{K}^{0}\phi$ etc.
For example, it is expected that Br$(\bar{B}^{0}_{d}\rightarrow
\bar{K}^{0}\phi)\sim 10^{-5}$ [14].
Certainly such quantitative calculations involve large uncertainties [15].
Note that the pure penguin-induced decay modes $\bar{B}^{0}\rightarrow
\bar{K}^{0}\phi$
and $\bar{B}^{0}_{s}\rightarrow K^{0}\phi$ are good candidates for probing
direct $CP$
violation in the neutral $B$-meson system. They are measurable in the
forthcoming $B$
factories. Finally, let us remark the importance to study baryonic $B$ decays.
We believe that a comparison
between the theoretical predictions and experimental measurements of two-body
baryonic
$B$ decays, once it is possible, should provide  valuable information about the
creation
processes of quark-antiquark pairs inside hadrons.

\vspace{0.3cm}

	To lowest-order weak interactions in the standard model, we have presented a
complete quark-diagram scheme for the two-body mesonic or baryonic decays of
$B$
mesons. Some new remarks are given on applications of this graph language to
specific processes.
We discuss some pure channels of mesonic $B$ decays in detail, which can be
used to
test the graph scheme given here. In view of the fact that two-body nonleptonic
$B$ decays
are supplying valuable opportunities for the study of flavour mixing and $CP$
violation,
much more attention is worth paying to them in both theory and experiments. \\

	The author would like to thank Professor H. Fritzsch for his warm hospitality
and encouragement.
He also acknowledges helpful discussions with Professors D. Du, A.
Khodjamirian, and D.D. Wu.
He is finally indebted to the Alexander von Humboldt Foundation for its
financial support.

\newpage

\renewcommand{\baselinestretch}{1.2}

\newpage
{}.

\vspace{0.5cm}

\small
\begin{center}
\begin{tabular}{c|llll}\hline\hline \\
Quark	 	& \multicolumn{4}{c}{Pure decay modes} \\ \\
graph(s)	& $B^{-}_{u}\rightarrow f$ $~$	& $\bar{B}^{0}_{d}\rightarrow f$ $~$
		& $\bar{B}^{0}_{s}\rightarrow f$ $~$	& $B^{-}_{c}\rightarrow f$ \\ \\ \hline
\\
1		& $D^{-}_{s}\pi^{0}$	& $D^{+}K^{-}$		& $D^{-}K^{+}$		& 	\\
		&			& $D^{-}_{s}\pi^{+}$	& $D^{+}_{s}\pi^{-}$	& \\ \\

$1^{'}$		& 			& $\bar{D}^{0}\bar{K}^{0}$	& $D^{0}K^{0}$		& $D^{-}_{s}\pi^{0}$
\\
		&			& $D^{0}\bar{K}^{0}$		& $\bar{D}^{0}K^{0}$	&	\\ \\
2		&			& $D^{+}_{s}K^{-}$	& $D^{-}\pi^{+}$	& 	\\
		&			& $D^{-}_{s}K^{+}$	& $D^{+}\pi^{-}$	&	\\
		&			&			& $D^{0}\pi^{0}$		\\
		&			&			& $\bar{D}^{0}\pi^{0}$	&	\\ \\
$2^{'}$		&			& $D^{0}\phi$		& $J/\psi \pi^{0}$	&	\\
		&			& $\bar{D}^{0}\phi$	&			&	\\ \\

3		& $D^{-}\bar{K}^{0}$	&			&	& $K^{-}K^{0}$	\\
		& $D^{-}_{s}K^{0}$	&			&	& $K^{-}\pi^{0}$	\\
		&			&			&	& $\bar{K}^{0}\pi^{-}$	\\
		&			&			&	& $\pi^{-}\pi^{0}$	\\ \\
$3^{'}$		& $D^{-}\phi$		&			&	& $\pi^{-} \phi$	\\ \\
4-$4^{'}$	& 			& $\bar{K}^{0}\phi$	& $K^{0}\phi$	&		\\ \\
5-$5^{'}$	&			& $\phi\phi$		&		&		\\  \\ \hline\hline
\end{tabular}
\end{center}

\vspace{1.cm}

	Table 1: Some two-body mesonic $B$ decays induced by a single quark diagram or
a couple ones. Here the
pseudoscalar mesons $\pi, K$, and $D$ can be replaced by their corresponding
vector (or axial-vector)
counterparts $\rho$ ($a_{1}$), $K^{*}$, and $D^{*}$.

\newpage

\renewcommand{\baselinestretch}{0.9}

\footnotesize
\begin{center}
\begin{tabular}{lllllll}\hline\hline \\
Pure	& $~~$	& CKM	& $~~~$	& Branching ratio	& $~~$	& Branching ratio \\
channel	&	& factor&	& (BSW model [6])	& $~~$	& (Experiments [1])	\\ \\ \hline
\\
$B^{-}_{u}\rightarrow D^{-}_{s}\pi^{0}$		&& $V_{ub}V^{*}_{cs}$ 	&& $2.7\times
10^{-5}$	&& $<2.1\times 10^{-4}$ \\
$B^{-}_{u}\rightarrow D^{-}_{s}\rho^{0}$	&& $V_{ub}V^{*}_{cs}$	&& $1.3\times
10^{-5}$	&& $<4\times 10^{-4}$ \\
$B^{-}_{u}\rightarrow D^{-}_{s}a^{0}_{1}$	&& $V_{ub}V^{*}_{cs}$	&& $1.1\times
10^{-5}$	&& $<2.3\times 10^{-3}$ \\
$B^{-}_{u}\rightarrow D^{-}_{s}\omega$		&& $V_{ub}V^{*}_{cs}$	&& $1.3\times
10^{-5}$	&& $<5\times 10^{-4}$ \\
$B^{-}_{u}\rightarrow D^{*-}_{s}\pi^{0}$	&& $V_{ub}V^{*}_{cs}$	&& $1.9\times
10^{-5}$	&& $<3.4\times 10^{-4}$ \\
$B^{-}_{u}\rightarrow D^{*-}_{s}\rho^{0}$	&& $V_{ub}V^{*}_{cs}$	&& $3.6\times
10^{-5}$	&& $<5\times 10^{-4}$ \\
$B^{-}_{u}\rightarrow D^{*-}_{s}a^{0}_{1}$	&& $V_{ub}V^{*}_{cs}$	&& $2.6\times
10^{-5}$	&& $<1.7\times 10^{-3}$ \\
$B^{-}_{u}\rightarrow D^{*-}_{s}\omega$		&& $V_{ub}V^{*}_{cs}$	&& $3.5\times
10^{-5}$	&& $<7\times 10^{-4}$ \\  \\
$\bar{B}^{0}_{d}\rightarrow D^{+}K^{-}$		&& $V_{cb}V^{*}_{us}$	&& $2.5\times
10^{-4}$	&& --- \\
$\bar{B}^{0}_{d}\rightarrow D^{+}K^{*-}$	&& $V_{cb}V^{*}_{us}$	&& $4.1\times
10^{-4}$	&& --- \\
$\bar{B}^{0}_{d}\rightarrow D^{*+}K^{-}$	&& $V_{cb}V^{*}_{us}$ 	&& $1.9\times
10^{-4}$  && --- \\
$\bar{B}^{0}_{d}\rightarrow D^{*+}K^{*-}$	&& $V_{cb}V^{*}_{us}$	&& $3.9\times
10^{-4}$	&& --- \\
$\bar{B}^{0}_{d}\rightarrow D^{-}_{s}\pi^{+}$	&& $V_{ub}V^{*}_{cs}$	&&
$5.6\times 10^{-5}$	&& $<2.9\times 10^{-4}$ \\
$\bar{B}^{0}_{d}\rightarrow D^{-}_{s}\rho^{+}$	&& $V_{ub}V^{*}_{cs}$	&&
$2.7\times 10^{-5}$ 	&& $<7\times 10^{-4}$ \\
$\bar{B}^{0}_{d}\rightarrow D^{-}_{s}a_{1}^{+}$	&& $V_{ub}V^{*}_{cs}$	&&
$2.1\times 10^{-5}$ 	&& $<2.7\times 10^{-3}$ \\
$\bar{B}^{0}_{d}\rightarrow D^{*-}_{s}\pi^{+}$	&& $V_{ub}V^{*}_{cs}$	&&
$7.2\times 10^{-5}$	&& $<5\times 10^{-4}$ \\
$\bar{B}^{0}_{d}\rightarrow D^{*-}_{s}\rho^{+}$	&& $V_{ub}V^{*}_{cs}$	&&
$1.4\times 10^{-4}$ 	&& $<2.2\times 10^{-4}$ \\
$\bar{B}^{0}_{d}\rightarrow D^{*-}_{s}a_{1}^{+}$&& $V_{ub}V^{*}_{cs}$	&&
$1.0\times 10^{-3}$	&& $<2.2\times 10^{-3}$ \\ \\
$\bar{B}^{0}_{s}\rightarrow D^{-}K^{+}$		&& $V_{ub}V^{*}_{cd}$	&& $1.6\times
10^{-6}$	&& --- \\
$\bar{B}^{0}_{s}\rightarrow D^{-}K^{*+}$	&& $V_{ub}V^{*}_{cd}$	&& $8.3\times
10^{-7}$	&& --- \\
$\bar{B}^{0}_{s}\rightarrow D^{*-}K^{+}$	&& $V_{ub}V^{*}_{cd}$	&& $1.5\times
10^{-6}$	&& --- \\
$\bar{B}^{0}_{s}\rightarrow D^{*-}K^{*+}$	&& $V_{ub}V^{*}_{cd}$	&& $1.9\times
10^{-6}$ 	&& --- \\
$\bar{B}^{0}_{s}\rightarrow D^{+}_{s}\pi^{-}$	&& $V_{cb}V^{*}_{ud}$	&&
$4.0\times 10^{-3}$	&& --- \\
$\bar{B}^{0}_{s}\rightarrow D^{+}_{s}\rho^{-}$	&& $V_{cb}V^{*}_{ud}$	&&
$1.0\times 10^{-2}$	&& --- \\
$\bar{B}^{0}_{s}\rightarrow D^{+}_{s}a_{1}^{-}$	&& $V_{cb}V^{*}_{ud}$	&&
$8.7\times 10^{-3}$	&& --- \\
$\bar{B}^{0}_{s}\rightarrow D^{*+}_{s}\pi^{-}$	&& $V_{cb}V^{*}_{ud}$	&&
$3.0\times 10^{-3}$	&& --- \\
$\bar{B}^{0}_{s}\rightarrow D^{*+}_{s}\rho^{-}$	&& $V_{cb}V^{*}_{ud}$	&&
$9.0\times 10^{-3}$	&& --- \\
$\bar{B}^{0}_{s}\rightarrow D^{*+}_{s}a_{1}^{-}$&& $V_{cb}V^{*}_{ud}$	&&
$9.6\times 10^{-3}$	&& --- \\ \\ \hline \\
$\bar{B}^{0}_{d}\rightarrow \bar{D}^{0}\bar{K}^{0}$	&& $V_{ub}V^{*}_{cs}$	&&
$2.1\times 10^{-6}$	&& --- \\
$\bar{B}^{0}_{d}\rightarrow \bar{D}^{0}\bar{K}^{*0}$	&& $V_{ub}V^{*}_{cs}$	&&
$1.0\times 10^{-6}$	&& --- \\
$\bar{B}^{0}_{d}\rightarrow \bar{D}^{*0}\bar{K}^{0}$	&& $V_{ub}V^{*}_{cs}$	&&
$1.5\times 10^{-6}$ 	&& --- \\
$\bar{B}^{0}_{d}\rightarrow \bar{D}^{*0}\bar{K}^{*0}$	&& $V_{ub}V^{*}_{cs}$	&&
$2.4\times 10^{-6}$	&& --- \\
$\bar{B}^{0}_{d}\rightarrow D^{0}\bar{K}^{0}$	&& $V_{cb}V^{*}_{us}$	&&
$1.3\times 10^{-5}$	&& --- \\
$\bar{B}^{0}_{d}\rightarrow D^{0}\bar{K}^{*0}$	&& $V_{cb}V^{*}_{us}$	&&
$6.5\times 10^{-6}$	&& --- \\
$\bar{B}^{0}_{d}\rightarrow D^{*0}\bar{K}^{0}$	&& $V_{cb}V^{*}_{us}$	&&
$9.3\times 10^{-6}$	&& --- \\
$\bar{B}^{0}_{d}\rightarrow D^{*0}\bar{K}^{*0}$	&& $V_{cb}V^{*}_{us}$	&&
$1.5\times 10^{-5}$	&& --- \\ \\

$\bar{B}^{0}_{s}\rightarrow D^{0}K^{0}$		&& $V_{cb}V^{*}_{ud}$	&& $3.2\times
10^{-4}$	&& --- \\
$\bar{B}^{0}_{s}\rightarrow D^{0}K^{*0}$	&& $V_{cb}V^{*}_{ud}$	&& $1.6\times
10^{-4}$	&& --- \\
$\bar{B}^{0}_{s}\rightarrow D^{*0}K^{0}$	&& $V_{cb}V^{*}_{ud}$	&& $2.3\times
10^{-4}$	&& --- \\
$\bar{B}^{0}_{s}\rightarrow D^{*0}K^{*0}$	&& $V_{cb}V^{*}_{ud}$	&& $3.8\times
10^{-4}$	&& --- \\
$\bar{B}^{0}_{s}\rightarrow \bar{D}^{0}K^{0}$	&& $V_{ub}V^{*}_{cd}$	&&
$1.2\times 10^{-7}$	&& --- \\
$\bar{B}^{0}_{s}\rightarrow \bar{D}^{0}K^{*0}$	&& $V_{ub}V^{*}_{cd}$	&&
$5.9\times 10^{-8}$	&& --- \\
$\bar{B}^{0}_{s}\rightarrow \bar{D}^{*0}K^{0}$	&& $V_{ub}V^{*}_{cd}$	&&
$8.5\times 10^{-8}$	&& --- \\
$\bar{B}^{0}_{s}\rightarrow \bar{D}^{*0}K^{*0}$	&& $V_{ub}V^{*}_{cd}$	&&
$1.4\times 10^{-7}$	&& --- \\ \\ \hline\hline
\end{tabular}
\end{center}

\vspace{1.cm}
\small

	Table 2: Predicted branching ratios for some two-body mesonic $B$ decays
induced by a single spectator (1 or $1^{'}$)
quark diagram in the context of the BSW model. We have used $|V_{cb}|\approx
0.04$ and $|V_{ub}|\approx 0.0035$.
The values of meson masses are quoted from Ref. [1], and the values of decay
constants and formfactors are
quoted from Refs. [1,6,10].

\newpage

\small
\begin{figure}
\begin{picture}(400,240)(-50,0)
\put(70,215){\line(1,0){90}}
\put(70,190){\line(1,0){90}}
\put(160,240){\oval(70,15)[l]}
\put(145,247.5){\vector(1,0){2}}
\put(145,232.5){\vector(-1,0){2}}
\put(85,215){\vector(1,0){2}}
\put(145,215){\vector(1,0){2}}
\put(85,190){\vector(-1,0){2}}
\put(145,190){\vector(-1,0){2}}
\multiput(110,215)(3,5){5}{\line(0,1){5}}
\multiput(107,215)(3,5){6}{\line(1,0){3}}
\put(113,165){(1)}
\put(250,240){\line(1,0){90}}
\put(250,190){\line(1,0){90}}
\put(340,215){\oval(70,25)[l]}
\put(265,240){\vector(1,0){2}}
\put(265,190){\vector(-1,0){2}}
\put(325,240){\vector(1,0){2}}
\put(325,190){\vector(-1,0){2}}
\put(325,227.5){\vector(-1,0){2}}
\put(325,202.5){\vector(1,0){2}}
\multiput(290,240)(3,-5){5}{\line(0,-1){5}}
\multiput(287,240)(3,-5){6}{\line(1,0){3}}
\put(293,165){($1'$)}
\put(70,114.5){\line(1,0){90}}
\put(70,65.5){\line(1,0){90}}
\put(160,90){\oval(80,25)[l]}
\put(85,114.5){\vector(1,0){2}}
\put(85,65.5){\vector(-1,0){2}}
\put(145,114.5){\vector(1,0){2}}
\put(145,65.5){\vector(-1,0){2}}
\put(145,102.5){\vector(-1,0){2}}
\put(145,77.5){\vector(1,0){2}}
\multiput(100,108.3)(0,-6){8}{$>$}
\put(113,40){(2)}
\put(250,90){\line(1,0){90}}
\put(250,65){\line(1,0){90}}
\put(340,115){\oval(70,15)[l]}
\put(265,90){\vector(1,0){2}}
\put(265,65){\vector(-1,0){2}}
\put(325,90){\vector(1,0){2}}
\put(325,65){\vector(-1,0){2}}
\put(325,122.5){\vector(1,0){2}}
\put(325,107.5){\vector(-1,0){2}}
\multiput(280,84)(0,-6){4}{$>$}
\put(293,40){($2'$)}
\put(70,-35){\oval(60,31)[r]}
\put(160,-35){\oval(52,21)[l]}
\put(160,-35){\oval(77,46)[l]}
\put(85,-19.5){\vector(1,0){2}}
\put(85,-50.0){\vector(-1,0){2}}
\put(145,-24.5){\vector(-1,0){2}}
\put(145,-45.5){\vector(1,0){2}}
\put(145,-12){\vector(1,0){2}}
\put(145,-58){\vector(-1,0){2}}
\multiput(99.5,-35)(5.2,0){4}{$\wedge$}
\put(113,-85){(3)}
\put(250,-44.5){\oval(60,31)[r]}
\put(340,-44.5){\oval(68,25)[l]}
\put(340,-10){\oval(68,15)[l]}
\put(265,-29){\vector(1,0){2}}
\put(265,-60){\vector(-1,0){2}}
\put(325,-2.5){\vector(1,0){2}}
\put(325,-17.5){\vector(-1,0){2}}
\put(325,-32){\vector(1,0){2}}
\put(325,-57){\vector(-1,0){2}}
\multiput(279.5,-44.5)(5.1,0){5}{$\wedge$}
\put(293,-85){($3'$)}
\end{picture}

\vspace{-2cm}
\begin{picture}(400,430)(-50,0)
\put(70,240){\line(1,0){25}}
\put(125,240){\line(1,0){35}}
\put(110,240){\oval(30,36)[b]}
\put(70,190){\line(1,0){90}}
\put(160,215){\oval(60,25)[l]}
\put(145,240){\vector(1,0){2}}
\put(145,227.5){\vector(-1,0){2}}
\put(145,202.5){\vector(1,0){2}}
\put(85,240){\vector(1,0){2}}
\put(85,190){\vector(-1,0){2}}
\put(145,190){\vector(-1,0){2}}
\multiput(94,240)(5,0){6}{$\wedge$}
\put(113,165){(4)}
\put(250,215){\line(1,0){30}}
\put(310,215){\line(1,0){30}}
\put(295,215){\oval(30,36)[b]}
\put(250,190){\line(1,0){90}}
\put(340,240){\oval(80,15)[l]}
\put(265,215){\vector(1,0){2}}
\put(265,190){\vector(-1,0){2}}
\put(325,215){\vector(1,0){2}}
\put(325,190){\vector(-1,0){2}}
\put(325,232.5){\vector(-1,0){2}}
\put(325,247.5){\vector(1,0){2}}
\multiput(279,215)(5,0){6}{$\wedge$}
\put(293,165){($4'$)}
\put(70,90){\oval(84,31)[r]}
\put(160,90){\oval(52,21)[l]}
\put(160,90){\oval(74,46)[l]}
\put(85,105.5){\vector(1,0){2}}
\put(85,74.5){\vector(-1,0){2}}
\put(145,113){\vector(1,0){2}}
\put(145,67){\vector(-1,0){2}}
\put(145,100.5){\vector(-1,0){2}}
\put(145,79.5){\vector(1,0){2}}
\multiput(92,99.5)(0,-6){5}{$>$}
\put(113,40){(5)}
\put(250,90){\oval(84,31)[r]}
\put(340,105){\oval(70,18)[l]}
\put(340,75){\oval(70,18)[l]}
\put(265,105.5){\vector(1,0){2}}
\put(265,74.5){\vector(-1,0){2}}
\put(325,114){\vector(1,0){2}}
\put(325,96){\vector(-1,0){2}}
\put(325,84){\vector(1,0){2}}
\put(325,66){\vector(-1,0){2}}
\multiput(272,99.5)(0,-6){5}{$>$}
\put(293,40){($5'$)}
\end{picture}
\vspace{-0.2cm}
\caption{Quark diagrams for two-body mesonic $B$ (or $D$) decays.}

\end{figure}

\newpage

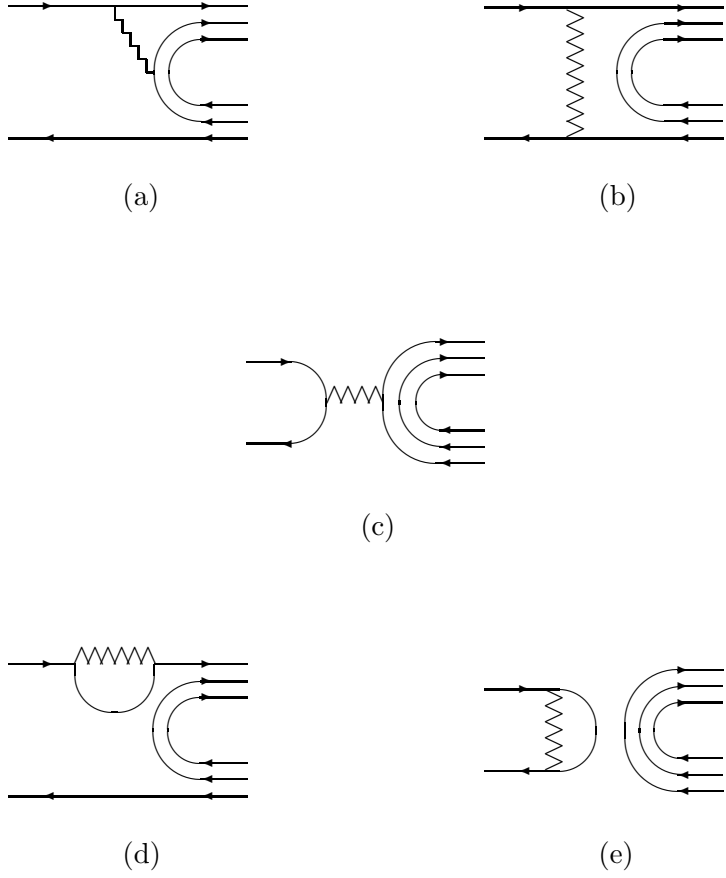
\begin{figure}
\begin{picture}(400,240)(-50,0)
\put(70,240){\line(1,0){90}}
\put(70,190){\line(1,0){90}}
\put(160,215){\oval(70,37.5)[l]}
\put(160,215){\oval(59,25)[l]}
\put(85,240){\vector(1,0){2}}
\put(85,190){\vector(-1,0){2}}
\put(145,240){\vector(1,0){2}}
\put(145,190){\vector(-1,0){2}}
\put(145,227.5){\vector(1,0){2}}
\put(145,202.5){\vector(-1,0){2}}
\put(145,196.25){\vector(-1,0){2}}
\put(145,233.75){\vector(1,0){2}}
\multiput(110,240)(3,-5){5}{\line(0,-1){5}}
\multiput(107,240)(3,-5){6}{\line(1,0){3}}
\put(113,165){(a)}
\put(250,239.5){\line(1,0){90}}
\put(250,190){\line(1,0){90}}
\put(340,215){\oval(80,37)[l]}
\put(340,215){\oval(69,25)[l]}
\put(265,239.5){\vector(1,0){2}}
\put(265,190){\vector(-1,0){2}}
\put(325,239.5){\vector(1,0){2}}
\put(325,190){\vector(-1,0){2}}
\put(325,227.5){\vector(1,0){2}}
\put(325,202.5){\vector(-1,0){2}}
\put(325,196.5){\vector(-1,0){2}}
\put(325,233.5){\vector(1,0){2}}
\multiput(280,232.8)(0,-6){8}{$>$}
\put(293,165){(b)}
\put(160,90){\oval(60,31)[r]}
\put(250,90){\oval(52,21)[l]}
\put(250,90){\oval(77,46)[l]}
\put(250,90){\oval(64.5,33.5)[l]}
\put(175,105.5){\vector(1,0){2}}
\put(175,74.5){\vector(-1,0){2}}
\put(235,100.5){\vector(1,0){2}}
\put(235,79.5){\vector(-1,0){2}}
\put(235,106.75){\vector(1,0){2}}
\put(235,73.25){\vector(-1,0){2}}
\put(235,113){\vector(1,0){2}}
\put(235,67){\vector(-1,0){2}}
\multiput(189.5,90)(5.2,0){4}{$\wedge$}
\put(203,40){(c)}
\end{picture}

\vspace{-2cm}
\begin{picture}(400,430)(-50,0)
\put(70,365){\line(1,0){25}}
\put(125,365){\line(1,0){35}}
\put(110,365){\oval(30,36)[b]}
\put(70,315){\line(1,0){90}}
\put(160,340){\oval(60,25)[l]}
\put(160,340){\oval(71,37)[l]}
\put(145,365){\vector(1,0){2}}
\put(145,352.5){\vector(1,0){2}}
\put(145,327.5){\vector(-1,0){2}}
\put(145,358.5){\vector(1,0){2}}
\put(145,321.5){\vector(-1,0){2}}
\put(85,365){\vector(1,0){2}}
\put(85,315){\vector(-1,0){2}}
\put(145,315){\vector(-1,0){2}}
\multiput(94,365)(5,0){6}{$\wedge$}
\put(113,290){(d)}
\put(250,340){\oval(84,31)[r]}
\put(340,340){\oval(52,21)[l]}
\put(340,340){\oval(63,33.5)[l]}
\put(340,340){\oval(74,46)[l]}
\put(265,355.5){\vector(1,0){2}}
\put(265,324.5){\vector(-1,0){2}}
\put(325,363){\vector(1,0){2}}
\put(325,317){\vector(-1,0){2}}
\put(325,350.5){\vector(1,0){2}}
\put(325,329.5){\vector(-1,0){2}}
\put(325,356.75){\vector(1,0){2}}
\put(325,323.25){\vector(-1,0){2}}
\multiput(272,349.5)(0,-6){5}{$>$}
\put(293,290){(e)}
\end{picture}
\vspace{-9cm}
\caption{Quark diagrams for two-body baryonic $B$ decays.}

\end{figure}

\begin{figure}
\begin{picture}(400,280)(-50,0)
\put(70,215){\line(1,0){90}}
\put(62,213){$c$}
\put(62,186){$\bar{b}$}
\put(43,200){$B^{+}_{c}$}
\put(163,212){$d,s$}
\put(163,186){$\bar{b}$}
\put(163,245){$\nu^{~}_{l}$}
\put(163,228){$l^{+}$}
\put(180,199){$B^{0}_{d,s}$}
\put(70,190){\line(1,0){90}}
\put(160,240){\oval(70,15)[l]}
\put(145,247.5){\vector(1,0){2}}
\put(145,232.5){\vector(-1,0){2}}
\put(85,215){\vector(1,0){2}}
\put(145,215){\vector(1,0){2}}
\put(85,190){\vector(-1,0){2}}
\put(145,190){\vector(-1,0){2}}
\multiput(110,215)(3,5){5}{\line(0,1){5}}
\multiput(107,215)(3,5){6}{\line(1,0){3}}
\put(113,165){(a)}
\put(270,240){\line(1,0){90}}
\put(270,190){\line(1,0){90}}
\put(262,238){$c$}
\put(262,186){$\bar{b}$}
\put(243,210.5){$B^{+}_{c}$}
\put(363,238){$s$}
\put(363,186){$\bar{b}$}
\put(363,223){$\bar{d}$}
\put(363,200){$u$}
\put(374,193){$B^{+}_{u}$}
\put(374,230){$\bar{K}^{0}$}
\put(360,215){\oval(70,25)[l]}
\put(285,240){\vector(1,0){2}}
\put(285,190){\vector(-1,0){2}}
\put(345,240){\vector(1,0){2}}
\put(345,190){\vector(-1,0){2}}
\put(345,227.5){\vector(-1,0){2}}
\put(345,202.5){\vector(1,0){2}}
\multiput(310,240)(3,-5){5}{\line(0,-1){5}}
\multiput(307,240)(3,-5){6}{\line(1,0){3}}
\put(313,165){(b)}
\end{picture}
\vspace{-4.8cm}
\caption{Examples for $B^{+}_{c}$ decays to $B^{+}_{u}$, $B^{0}_{d}$, and
$B^{0}_{s}$ mesons.}
\end{figure}
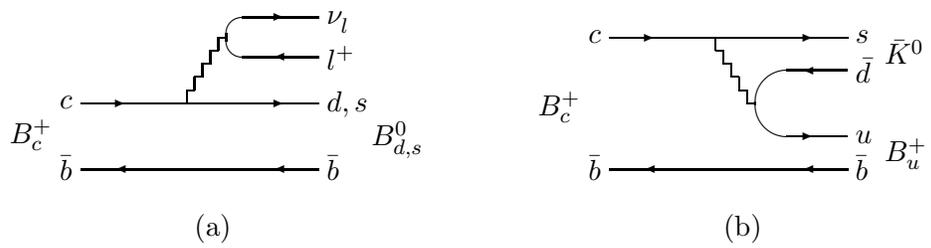

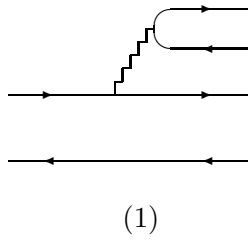
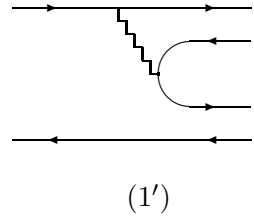
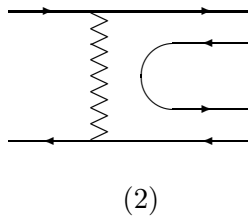
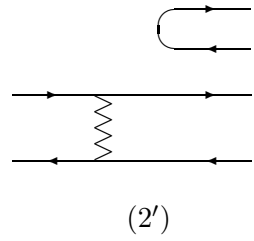
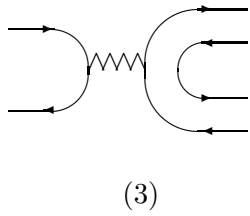
\end{document}